\documentclass[sigconf]{acmart}

\usepackage{amsmath}
\usepackage{amssymb}
\usepackage{graphicx}
\usepackage{textcomp}
\usepackage{xcolor}

\usepackage{booktabs} % For formal tables
\usepackage{float}
\usepackage{multirow}
\usepackage{indentfirst}
\usepackage{multirow}
\usepackage{graphicx}
\usepackage{subfigure}
\usepackage[]{graphicx,indentfirst}
\usepackage{arydshln}
\usepackage{textcomp}
\usepackage{multicol}
\usepackage{bm}
\usepackage{url}
\usepackage{amsfonts}
\usepackage{mathrsfs}
\usepackage[ruled]{algorithm2e}

\AtBeginDocument{%
  \providecommand\BibTeX{{%
    \normalfont B\kern-0.5em{\scshape i\kern-0.25em b}\kern-0.8em\TeX}}}

\copyrightyear{2023} 
\acmYear{2023} 
\setcopyright{acmcopyright}
\acmConference[WSDM '23]{Proceedings of the Sixteenth ACM International Conference on Web Search and Data Mining}{February 27-March 3, 2023}{Singapore, Singapore}
\acmBooktitle{Proceedings of the Sixteenth ACM International Conference on Web Search and Data Mining (WSDM '23), February 27-March 3, 2023, Singapore, Singapore}
\acmPrice{15.00}
\acmDOI{10.1145/3539597.3570482}
\acmISBN{978-1-4503-9407-9/23/02}

\begin{document}

\title{VRKG4Rec: Virtual Relational Knowledge Graph for Recommendation}

%\email{webmaster@marysville-ohio.com}
%\affiliation{%
%  \institution{Institute for Clarity in Documentation}
%  \streetaddress{P.O. Box 1212}
%  \country{USA}
\author{Lingyun Lu}
\affiliation{%
	\institution{School of Electronic Information and Communications, Huazhong University of Science and Technology}
	\streetaddress{Huazhong University of Science and Technology}
	\city{Wuhan}
	\country{China}}
\email{lulingyun@hust.edu.cn}

\author{Bang Wang}
\affiliation{%
	\institution{School of Electronic Information and Communications, Huazhong University of Science and Technology}
	\streetaddress{Huazhong University of Science and Technology}
	\city{Wuhan}
	\country{China}}
\email{wangbang@hust.edu.cn}

\author{Zizhuo Zhang}
\affiliation{%
	\institution{School of Electronic Information and Communications, Huazhong University of Science and Technology}
	\streetaddress{Huazhong University of Science and Technology}
	\city{Wuhan}
	\country{China}}
\email{zhangzizhuo@hust.edu.cn}

\author{Shenghao Liu}
\affiliation{%
	\institution{School of Cyber Science and Engineering, Huazhong University of Science and Technology}
	\streetaddress{Huazhong University of Science and Technology}
	\city{Wuhan}
	\country{China}}
\email{liushenghao@hust.edu.cn}

\author{Han Xu}
\affiliation{%
	\institution{School of Journalism and Information Communication, Huazhong University of Science and Technology}
	\streetaddress{Huazhong University of Science and Technology}
	\city{Wuhan}
	\country{China}}
\email{xuh@hust.edu.cn}

%\affiliation{%
%  \institution{Inria Paris-Rocquencourt}
%\affiliation{%
% \institution{Rajiv Gandhi University}
%\affiliation{%
%  \institution{Tsinghua University}
%  \streetaddress{30 Shuangqing Rd}
%\affiliation{%
%  \institution{Palmer Research Laboratories}
%  \streetaddress{8600 Datapoint Drive}
%  \country{USA}
%\affiliation{%
%  \institution{The Th{\o}rv{\"a}ld Group}
%  \streetaddress{1 Th{\o}rv{\"a}ld Circle}
%\email{jsmith@affiliation.org}
%\affiliation{%
%  \institution{The Kumquat Consortium}
%  \city{New York}
%\email{jpkumquat@consortium.net}

\renewcommand{\shortauthors}{Lu et al.}

\begin{abstract}
Incorporating knowledge graph as side information has become a new trend in recommendation systems. Recent studies regard items as entities of a knowledge graph and leverage graph neural networks to assist item encoding, yet by considering each relation type independently. However, relation types are often too many and sometimes one relation type involves too few entities. We argue that there may exist some latent relevance among relations in KG. It may not necessary nor effective to consider all relation types for item encoding. In this paper, we propose a VRKG4Rec model (\underline{V}irtual \underline{R}elational \underline{K}nowledge \underline{G}raphs \underline{f}or \underline{Rec}ommendation), which clusters relations with latent relevance to generates virtual relations. Specifically, we first construct virtual relational graphs (VRKGs) by an unsupervised learning scheme. We also design a \textit{local weighted smoothing} (LWS) mechanism for node encoding on VRKGs, which iteratively updates a node embedding only depending on the node itself and its neighbors, but involve no additional training parameters. LWS mechanism is also employed on a user-item bipartite graph for user representation learning, which utilizes item encodings with virtual relational knowledge to help train user representations. Experiment results on two public datasets validate that our VRKG4Rec model outperforms the state-of-the-art methods. The implementations are available at https://github.com/lulu0913/VRKG4Rec.
\end{abstract}





\maketitle
\section{Introduction}\label{sec:introduction}
Recommender system (RS)~\cite{guo:2020:IEEE:survey, wu:2020:CSUR:survey, zhang:2020:Neurocomputing} as an information filtering tool plays an important role in our daily life to deal with the problem of information overload~\cite{Chen:2021:WSDM}. Traditional RS mainly depends on collaborative filtering of user-item interaction records, which suffers from the data sparsity problem ~\cite{zhang:2021:InfoSci, liu:2021:hnf}. Recently, knowledge graph (KG)~\cite{wang:2017:IEEE:knowledgesurvey} with entities and rich relation connections has been exploited in RS to serve as side information for alleviating this problem~\cite{zou:2022:SIGIR, zou:2022:CIKM}. The entities in KG can describe the attributes of the item, thus enhancing the item content representation. Generally speaking, a KG consists of many knowledge triplets $(h,r,t)$, where $h$ and $t$ are a head entity and a tail entity respectively, and $r$ is the relation from $h$ to $t$. For example, (\textit{Leonardo}, \textit{star}, \textit{Titanic}) states the fact that Leonardo is the star of the movie Titanic. A KG can contain multiple types of relations, and more than one type of relation can exist between $h$ and $t$ in a KG. This makes it possible for KG to transfer the complex relationships in real world.

The main challenge of KG-based recommendation is how to extract useful information from structured knowledge triplets for better learning item representation. Some researches ~\cite{zhang:2016:KDD:CKE,ai:2018:algorithms:ECFKG,wang:2019:WWW:MKR} first learn entity representation in a KG by some knowledge representation learning method (e.g. the TransX series~\cite{bordes:2013:transE, yankai:2015:AAAI:transR, wang:2016:transH, ji:2015:transD}) and use the learned embeddings as pre-trained features in a follow-up recommendation task. Although these methods can well model knowledge triplet relationships in a KG, they mainly focus on the first-order connectivity but ignore the global relationships of entities. Some researches ~\cite{catherine:2016:recsys,ma:2019:WWW} explore a kind of meta-path connecting history entity and the candidate entity. However, these methods suffer from two main problems: (1) Devising such meta-path requires strong expertise, which has poor transfer capability. (2) It is labor-intensive and time-consuming when applying on the KGs of large scale.

Recently, some researches~\cite{wang:2019:KDD:KGNN-LS,Wang:2019:KDD:KGAT, wang:2020:SIGIR:CKAN, wang:2018:WWW:DKN} use graph neural networks (GNNs) for encoding knowledge in terms of entities and their relations into item representation and have achieved excellent and promising performance in recommendation. The key idea of these GNN-based models is to aggregate the information from direct-neighbor into node representation and incorporate the information from high-order neighbors by stacking multiple GNN layers. ~\cite{zhang:2021:ICDM, zhang:2022:CIKM} Despite their significant performance improvements, we argue that they still suffer from a common problem. 

\begin{figure}[t]
	\centering
	\includegraphics[width=0.47\textwidth]{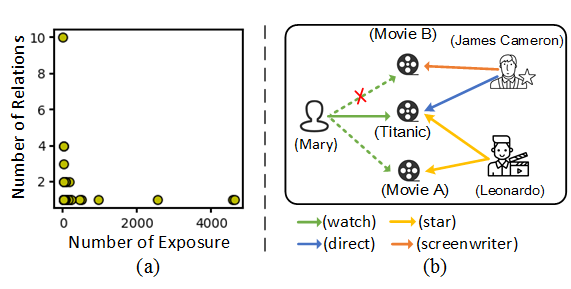}
	\caption{Illustration of two kinds of motivations. (a)The long-tail relation distribution of Last.FM dataset. (b)An illustration example of necessity of considering the relevance of different relations}
	\label{fig:intro}
\end{figure}

Existing methods~\cite{wang:2019:WWW:KGCN, wang:2019:KDD:KGNN-LS, wang:2021:WWW:KGIN, wang:2018:CIKM:ripplenet, Wang:2019:KDD:KGAT, wang:2020:SIGIR:CKAN} apply the devised aggregation mechanism directly on the input KG which means KG quality is crucial for a distinguished performance. However, in real-world dataset, the relations in KG always exhibit long-tail distribution. The distribution of KG relations collected from real-world dataset Last-FM\footnote{https://grouplens.org/datasets/hetrec-2011/} is reported in Fig.~\ref{fig:intro}(a), where x-axis is the relation exposure count in KG and y-axis represents the corresponding number of relations. From Fig.~\ref{fig:intro}(a), we notice that the relations in KG show long-tail issue, which means most of the relations appear only a few times. This will affect user preference propagation in KG if we treat these relations as totally irrelevant types of edges. For example, in Fig.~\ref{fig:intro}(b), there are four triplets in KG. They are \textit{triplet 1=(James Cameron, screenwriter, Movie B)}, \textit{triplet 2=(James Cameron, direct, Titanic)}, \textit{triplet 3=(Leonardo, star, Titanic)} and \textit{triplet 4=(Leonardo, star, Movie A)}. It can be deduced from \textit{triplet 3} and \textit{triplet 4} that \textit{Movie A} will be recommended to Mary since Mary has watched \textit{Titanic} and it has the same movie star \textit{Leonardo}, while \textit{Movie B} may not recommended to Mary for \textit{triplet 1} and \textit{triplet 2} have different relations, that is \textit{direct} and \textit{screenwriter}. However, \textit{direct} and \textit{screenwriter} have the similar semantic. So Mary will also interested in \textit{Movie B}. As a result, it is not wise to consider every relation independently, especially in long-tail situation that most relations only exist in a few triplets.

%Besides, rich entities as KG has, it may be sparse when only considering part of the graph consisting of only one type of relation. 

To deal with the problem, in this paper, we propose to generate virtual relations for KG and design a \underline{V}irtual \underline{R}elational \underline{K}nowledge \underline{G}raph \underline{f}or \underline{Rec}ommendation model (VRKG4Rec). We first construct \textit{virtual relational knowledge graphs} (VRKGs) from an unsupervised learning approach. Given entities contribute differently to item, we next devise a local weighted smoothing (LWS) mechanism on each VRKG to learn aggregated relational knowledge for item encoding while keeping the semantic independence. We obtain the final representation by fusing item encodings from all VRKGs. We also employ the LWS mechanism on a user-item bipartite graph to learn user representation, which utilizes item encodings with relational knowledge to help train user representations. Experiment results on two public datasets validate that our VRKG4Rec model outperforms the state-of-the-art methods.

we summarize the contribution of this work as follows:
\begin{itemize}
	\item Propose to construct Virtual Relational Knowledge Graph (VRKG) through unsupervised learning to reveal the relevant relations in item encoding and alleviate long-tail issues for recommendation.
	\item Devise a new aggregation scheme called Local Weighted Smoothing (LWS) mechanism to aggregate weighted neighbor knowledge into item representation.
	\item Conduct empirical studies on two benchmark datasets to demonstrate the proposed algorithm outperforms the state-of-the-art methods.
\end{itemize}
\par
The rest of the paper is structured as follows: Section~\ref{sec:Problem Formulation} gives the definition of the problem in this paper. The proposed VRKG4Rec model is presented in Section~\ref{sec:proposed model} and experimented in Section~\ref{sec:experiment}. We briefly reviews the related work in Section~\ref{sec:related work}. Section~\ref{sec:conlusion} concludes the paper with some discussions.

\begin{figure*}[t]
	\centering
	\includegraphics[width=0.95\textwidth, height=0.35\textwidth]{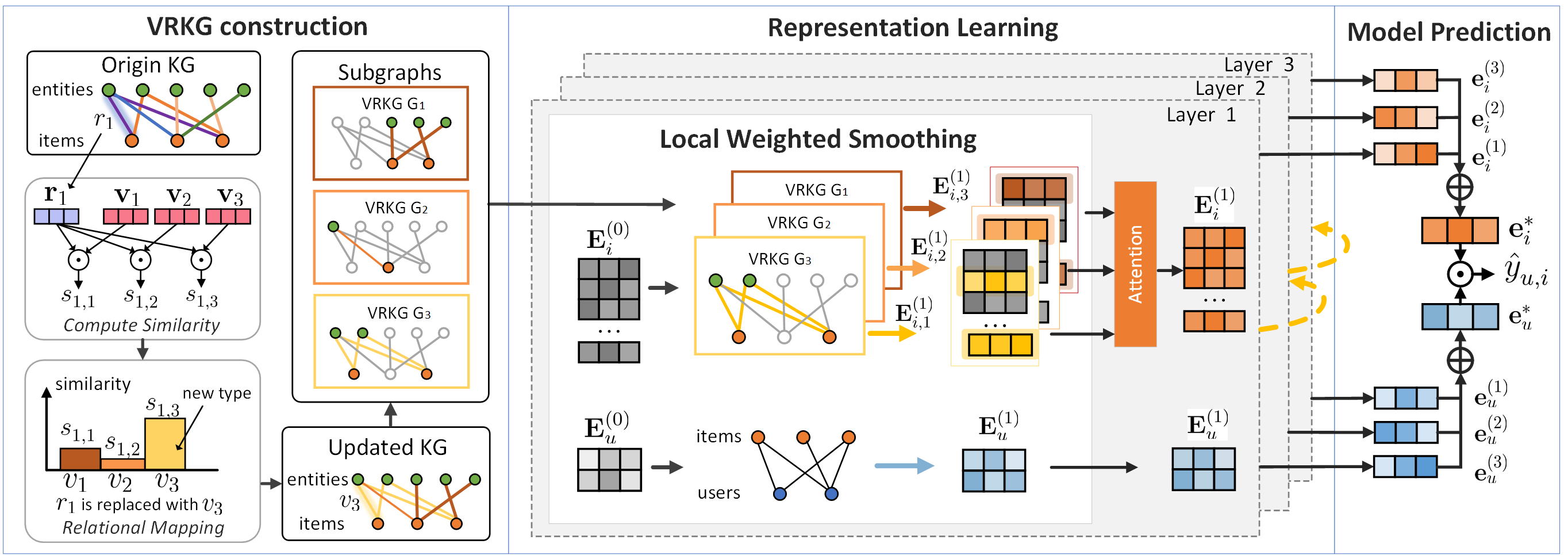}
	\caption{Overview of the proposed VRKG4Rec model}
	\label{fig:overall_model}
\end{figure*}

\section{Problem Formulation}\label{sec:Problem Formulation}
Knowledge-aware recommendation task is formulated as follows. The inputs of the recommendation model are interaction data in between users and items and a knowledge graph; while the output is a similarity score of a target user and a candidate item.
\par
The \textbf{interaction data} indicates users' interactions with items. There are total $M$ number of users and $N$ number of items. Let $\mathcal{U}={u_1,u_2,\cdots,u_M}$ and $\mathcal{I}={i_1,i_2,\cdots,i_N}$ be a set of users and a set of items, respectively. Let $O^{+}=\{(u, i) \mid u \in \mathcal{U}, i \in \mathcal{I}\}$ denote a set of positive interactions. Following the previous study KGAT~\cite{Wang:2019:KDD:KGAT}, we view the interaction as relation $ r^+ $ and we can construct user-item bipartite graph $ \mathcal{G}^I=\{(u, r^+, i)|(u, i)\in O^{+}\} $. Note that an item $ i $ in $ \mathcal{G}^I $ can be aligned with an entity $ h $ in knowledge graph.

The \textbf{knowledge graph} consists of many triplets which states real-world facts, such as the attributes of items or the relationship between entities. We use $\mathcal{G}=\{(h, r, t) \mid h, t \in \mathcal{E}, r \in \mathcal{R}\}$ to denote knowledge graph, where $\mathcal{E}$ is the set of entities and $\mathcal{R}$ is the relation set, which contains both canonical directions \textit{(e.g. Star)} and the inverse directions \textit{(e.g. StaredBy)} ; $ h, r $ and $ t $ is the head entity, relation and tail entity of the knowledge triplet respectively. In recommendation task, we align the items with entities $(\mathcal{I} \subset \mathcal{E})$ in knowledge graph so as to utilize the side information to facilitate item representation.

\section{The Proposed Model}\label{sec:proposed model}
In this section, we illustrate our VRKG4Rec model in detail and present the overview of the model in Fig.~\ref{fig:overall_model}. The proposed model consists of three key components: (1) Virtual Relational Knowledge Graph (VRKG) construction, which clusters all the relations in origin KG into $ K $ virtual relations by unsupervised learning and construct $K$ VRKGs. (2) Local Weighted Smoothing (LWS) aggregation module, which learns the representation of a node by weighted smoothing the representation of neighbor nodes to makes its representation closer to the most similar neighbors. (3) Representation learning module, which applies LWS to generate use and item representation. 

\subsection{VRKG Construction}\label{sec:VRKG_construction}
In this section, we first reveal the relevance relations for item encoding by clustering all types of original relations into $K$ virtual relations. Then we construct virtual relational knowledge graphs based on these virtual relations to demonstrate item in terms of different attributes. 

For the purpose of clustering relevant relations, we propose an unsupervised learning method to explore the latent factor of each original relation and fuse the original relations with similar latent factor into one kind of virtual relation. Firstly, we initialize the representations of such virtual relations as a virtual centroid matrix $\mathbf{V}\in \mathbb{R}^{K\times d}$
\begin{equation}
	\mathbf{V} = (\mathbf{v_1}, \mathbf{v_2}, ..., \mathbf{v_K})^{\mathsf{T}},
\end{equation}
where the $k$-th row $\mathbf{v}_k \in \mathbb{R}^d$ is the representation of the $k$-th virtual relation. $ K $ is the number of virtual relations and is a hyper-parameter which is set as 3 by default.
\par
On basis of this, we compute the similarities between the original relations in KG with the virtual relations. For the relation $r_p \in \mathcal{R}$, we construct its similarity vector $\mathbf{s}_p \in \mathbb{R}^{K}$ by
\begin{align}
	&\mathbf{s_p} = (g(\mathbf{r_p}, \mathbf{v_1}), g(\mathbf{r_p}, \mathbf{v_2}), ..., g(\mathbf{r_p}, \mathbf{v_K}))\\
	&g(\mathbf{r_p}, \mathbf{v_k}) = \mathbf{r_p}^{\top} \mathbf{v_k}
\end{align}
where $\mathbf{r_p} \in \mathbb{R}^{d}$ is the embedding of relation $r_p$, and $g(\cdot)$ is a similarity function, here we use inner product function.
\par
The relation $ r_p $ is replaced with a virtual relation $v_{k^{\prime}}$ with the highest similarity; While the knowledge graph $ \mathcal{G} $ will be updated as $ \mathcal{G}^{\prime} $accordingly.
\begin{align}
	k^{\prime} &=\arg\max \mathbf{s_p} \\
	&=\mathop{\arg\max}\limits_{k=1,2,...,K}(g(\mathbf{r_p}, \mathbf{v_1}),...,  g(\mathbf{r_p}, \mathbf{v_k}), ...) \\
	& (h, r_p, t) \leftarrow (h, v_{k^{\prime}}, t).
\end{align}
\par
The virtual centroid matrix $\mathbf{V}$ is trained together with the training of our recommendation model parameters $\Theta$. We note that such a joint training of $\mathbf{V}$ and $\Theta$ can help construct virtual relations more suitable for a downstream recommendation task.
\par
After model training, we divide the origin KG $ \mathcal{G} $ into some subgraphs, called virtual relational knowledge graphs (VRKGs), to capture the connection information between relations and preserve the semantic independence of item attributes. Specifically, VRKGs are constructed based on the updated knowledge graph $ \mathcal{G}^{\prime} $, each of which contains one of virtual relation and its connected entities from the knowledge graph$ \mathcal{G}^{\prime} $.
\begin{equation}
	\mathcal{G}_k = \{(h,r^{\prime} ,t)\mid (h,r^{\prime} ,t)\in \mathcal{G}^{\prime} , r^{\prime} =v_k\}.
\end{equation}

\subsection{Local Weighted Smoothing}\label{sec:LWS}
\begin{figure}[htp]
	\centering
	\includegraphics[width=0.45\textwidth]{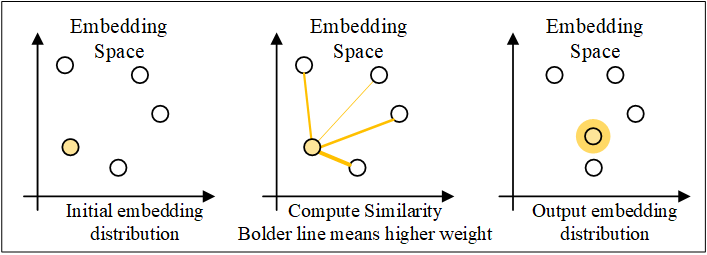}
	\caption{Core operation of LWS with single iteration}
	\label{fig:lws}
\end{figure}

For obtained VRKGs, We design a \textit{Local Weighted Smoothing} (LWS) mechanism to encode local information from direct neighbors into node embedding. The basic idea is to iteratively update the vector of a node closer to its direct neighbors that are similar in the embedding space and optimize the item embeddings to make them most advantageous for recommendation. The upper right part of Fig.~\ref{fig:lws} shows the core operation of the LWS mechanism with single iteration.

%To this end, the aggregation weight should be proportional to the embedding similarity. So LWS didn't introduce additional training parameters but use dot product similarity as aggregation weight and find out crucial entities for recommendation task in training process.

For an entity $h$ in $ \mathcal{G}_k $, the neighbor node set of $ h $ is denoted as $\mathcal{N}_k(h) = \{t | (h, v_k, t) \in \mathcal{G}_k \}$. To characterize the topology structure of VRKG $ \mathcal{G}_k $, we proposed to weighted smooth the embedding of neighbor nodes over the propagation:
\begin{eqnarray}
	\mathbf{e}_{\mathcal{N}_k(h)}^{(0)} = \sum_{t\in \mathcal{N}_k(h)} \pi (h, t)\mathbf{e}_t^{(0)},
\end{eqnarray}
where $ \mathbf{e}_t^{(0)} $ is the ID embedding of entity $ t $, $ \pi(h,t) $ is the weight of $ t $ in local information propagation. The weight reflects the contribution of entity $ t $ to depicting the attributes of entity $ h $. We argue that the more important entity $ t $ is, the more information should be propagated and entity $ h $ need to be closer to $ t $ in embedding space. For this purpose, we compute the similarity between entity $ t $ and $ h $ to serve as the weight $ \pi(h,t) $. Here, we use inner product to measure the similarity for simplicity:
\begin{eqnarray}
	\pi(h, t) = \mathbf{e}_{h}^{(0)\mathsf{T}}\cdot \mathbf{e}_{t}^{(0)},
\end{eqnarray}
where $ \mathbf{e}_h^{(0)} $ is the ID embedding of entity $ h $.

Then we aggregate the entity representation $ \mathbf{e}_h $ and the local information representation $ \mathbf{e}_{\mathcal{N}_k(h)} $ to obtain a temp representation as $ \mathbf{u}_h^{(1)} $:
\begin{align}
	\mathbf{u}_h^{(1)} &= {\rm AGG}( \mathbf{e}_h^{(0)}, \mathbf{e}_{\mathcal{N}_k(h)}^{(0)})\\
	&={\rm NORM}(\mathbf{e}_h^{(0)} + \mathbf{e}_{\mathcal{N}_k(h)}^{(0)}),
\end{align}
\par
We simply sum two representations up to update the entity representation without introduce additional training parameters in aggregator. Note that we didn't normalize the weight by adopting the softmax function in aggregation step like most knowledge-aware graph neural networks (e.g. KGAT~\cite{Wang:2019:KDD:KGAT}, KGCN~\cite{wang:2019:WWW:KGCN}) do. In this way, we can simplify the calculation. More importantly, it frees us from sampling a fixed number of neighbors for each node due to the limit of softmax computing and this enables us to take full advantage of knowledge graph without information loss in sampling. However, skipping normalization step may lead to the scale of the updated vector getting larger and becoming incomparable. To solve this problem, we normalize the temp vector $ \mathbf{u}_h^{(1)} $ by:
\begin{equation}
	{\rm NORM}(\mathbf{u}) = \frac{\mathbf{\mathbf{u}}}{\lVert \mathbf{\mathbf{u}}\rVert}\cdot \frac{\lVert \mathbf{\mathbf{u}} \rVert^{2}}{\lVert \mathbf{\mathbf{u}} \rVert^{2} +1},
\end{equation}
where $ ||\mathbf{u}|| $ is 2-norm of vector $ \mathbf{u} $. In this way, we can norm the scale of the vector to the range from 0 to 1, keeping its direction and relative sizes of vectors. We use $f_{agg}(\cdot)$  to denote the aggregation function.
\begin{equation}
	\mathbf{u}_{h,k}^{(1)} = f_{agg}\left( \{(\mathbf{e}_{h}^{(0)},\mathbf{e}_{t}^{(0)}) \mid t\in \mathcal{N}_{k}(h)\}\right)
\end{equation}

To further modify the representation of the node and make the distance in embedding space better reflects the similarity, we iteratively smooth the node vector based on the temp representation $ \mathbf{u}_h^{(1)} $ by repeat the above weighed propagation step:
\begin{equation}
	\mathbf{u}_{h,k}^{(q)} = f_{agg}\left( \{(\mathbf{u}_{h,k}^{(q-1)},\mathbf{e}_{t}^{(0)}) \mid t\in \mathcal{N}_{k}(h)\}\right),
\end{equation}
where $ q = 2, 3,...,Q $. We set $ Q = 3 $ as default in our experiments.
\par
Up to now, we finally update the representation of entity $ h $ as $ \mathbf{e}_{h,k}^{(1)} = \mathbf{u}_{h,k}^{(Q)} $. Note that during each iteration, we only smooth the vector of $ h $, keeping the neighbor vector unchanged. We use $f_{LWS}(\cdot)$ to denote the entire smooth operation of incorporating the first-order information into node embedding as follows:
\begin{align}
	&\mathbf{e}_{h,k}^{(1)} = f_{LWS}\left( \{(\mathbf{e}_{h}^{(0)},\mathbf{e}_{t}^{(0)}) \mid t\in \mathcal{N}_{k}(h)\}\right),
\end{align}
where $ \mathbf{e}_{h,k}^{(1)} $ is the updated representation of entity $ h $ in VKRG $ \mathcal{G}_k $. So far, we are able to encode the first-order information supplied by $ \mathcal{G}_k $ into item representation, forcing it closer to similar entity in embedding space.
%It is worth mentioning that the LWS computes the aggregation weights depending on node embedding themselves without trainable parameters.

\subsection{Representation Learning}
\subsubsection{Item Representation Learning.}
To preserve the semantic independence of the item attributes, we apply LWS on each of the $ K $ VRKGs respectively to encode item from K different aspects and then fuse these $ K $ embeddings by attention mechanism to acquire final item representation as  $ \mathbf{e}_h^{(1)} $:
\begin{align}
	\mathbf{e}_{h}^{(1)} &= \alpha_1 \mathbf{e}_{h,1}^{(1)}+\alpha_2 \mathbf{e}_{h,2}^{(1)}+...+\alpha_K \mathbf{e}_{h,K}^{(1)}\\
	\mathbf{e}_{h}^{(1)} &=\sum_{k=1}^{K} \alpha_k f_{LWS}\left( \{(\mathbf{e}_{h}^{(0)},\mathbf{e}_{t}^{(0)}) \mid t\in \mathcal{N}_{k}(h)\}\right),
\end{align}
where $ \mathbf{e}_{h}^{(1)} \in \mathbb{R}^d$ is the representation that collects the first-order connectivity information from knowledge graph. $ \alpha_k $ is the weight of the representation learned from VRKG $ \mathcal{G}_k $. It denotes the importance of this virtual relation in item encoding and is learned during training process.

In order to incorporate the information from high-order neighbors, we further stack more LWS aggregation layers. More formally, we recursively formulate the representation of entity $ h $ after $ l $ layers as:
\begin{equation}
	\mathbf{e}_{h}^{(l)} =\sum_{k=1}^{K} \alpha_k f_{LWS}\left( \{(\mathbf{e}_{h}^{(l-1)},\mathbf{e}_{t}^{(l-1)}) \mid t\in \mathcal{N}_{k}(h)\}\right),
\end{equation}
where $ \mathbf{e}_{h}^{(l)} $ denotes the representation of entity $ h $, which contains the information from its $ l $-hop neighbors, and $ l = 1, 2, ..., L $. We set $ L= 3 $ as default in our experiment.

\subsubsection{User Representation Learning}
To obtain user representation, we first construct a user-item bipartite graph based on the interaction data, which is denoted as $ \mathcal{G}^I=\{(u, r^+, i)|(u, i)\in O^{+}\} $. Specifically, there is an edge in between $u$ and $i$, only when the pair $(u,i)\in O^{+}$.
\par
Then, in order to make use of collaborative information and discriminate the impact of items that user has interacted before, we apply LWS on $ \mathcal{G}^I $ to unequally encode the information from historical interactions into user representation:
\begin{eqnarray}
	&&\mathbf{e}_{u}^{(l)} = f_{LWS}\left( \{(\mathbf{e}_{u}^{(l-1)},\mathbf{e}_{i}^{(l-1)}) \mid i\in \mathcal{N}(u)\}\right),
\end{eqnarray}
where $ l = 1,2,...,L $, $ \mathcal{N}(u) $ is a set of items that a user $ u $ has interacted with, $ \mathbf{e}_u^{(0)} $ and $ \mathbf{e}_i^{(0)} $ denote the ID embedding of the user $ u $ and item $ i $ respectively. Note that item $ i $ is aligned with entity $ h $ in KG and they have the same embedding, that is $ \mathbf{e}_i^{(l)} = \mathbf{e}_h^{(l)} $, where $ l = 1,2,...,L $. 

It worth mentioning that item nodes that both appear in the knowledge graph $ \mathcal{G} $ and the bipartite graph $ \mathcal{G}^I $ serve as a bridge to combine the two graphs. This allows information to be passed between graphs seamlessly. We propagate information on both graphs in parallel and train user and item representation simultaneously. By stacking multiple LWS layers, high-order relational knowledge from items can be propagated to user representation towards better recommendation performance.  
%Algorithm~\ref{Alg:LWS} describes the detailed LWS operations for learning user and item representation. We take user $ u $ and entity $ h $ (aligned with item $ i $) as an example.

\subsection{Prediction}
After $ L $ LWS layers, we obtain the representations of user $u$ and item $i$ (corresponding to node $ h $ in knowledge graph) at different layers. Given each output embedding focuses on the neighbors of different hops, we sum up the output of each layer as the final representation.
\begin{align}
	\mathbf{e}_{u}^{*} = \mathbf{e}_{u}^{(0)} + ... + \mathbf{e}_{u}^{(L)}\\
	\mathbf{e}_{i}^{*} = \mathbf{e}_{i}^{(0)} + ... + \mathbf{e}_{i}^{(L)}
\end{align}
Finally, we employ the inner product similarity for prediction as follows:
\begin{equation}
	\hat{y}_{ui} = \mathbf{e}_u^{(*)\top }\mathbf{e}_{i}^{(*)}.
\end{equation}

\subsection{Optimization}
To train this model, we construct the BPR loss function
\begin{equation}
	\mathcal{L}_{BPR} = \sum_{(u,i,j)\in \mathcal{O}}-\ln\sigma(\hat{y}_{ui} - \hat{y}_{uj}),
\end{equation}
where $\mathcal{O} = \{ (u,i,j)\mid (u,i)\in \mathcal{O}^+ , (u,j)\in \mathcal{O}^- \}$ is the training set. $ \mathcal{O}^+ $ is the set of positive interaction pairs which consists of historical user-item interactions; While $ \mathcal{O}^- $ is the set of negative pairs $ (u,j) $, where $ j $ is randomly sampled from items that user $ u $ has never interacted with. $\sigma(\cdot)$ is the sigmoid function.

The objective function for training model parameters is:
\begin{equation}
	\mathcal{L} = \mathcal{L}_{BPR} + \lambda {\lVert \Theta \rVert }_2^2,
\end{equation}
where $ \Theta = \{ \mathbf{e}_u^{(0)}, \mathbf{e}_i^{(0)}, \mathbf{r}_{p}, V, R \mid u\in \mathcal{U}, i \in \mathcal{E}, p\in \mathcal{R} \} $ is the model parameter set. $ \mathcal{E} $ is the entity set containing item set $ \mathcal{I} $, and generally $\mathcal{I}\subseteq \mathcal{E}$. $ \lambda $ is a hyperparameter controlling $ L_2 $ regularization to prevent overfitting.

\section{Experiments}\label{sec:experiment}
\begin{table}[h]
	\caption{Dataset Statistics}
	\label{table:dataset parameter}
    \resizebox{0.45\textwidth}{!}{
	\begin{tabular}{ c|ccc|ccc }
		\hline
		\multirow{2}{*}{dataset} 	&\multicolumn{3}{c|}{User-Item Interaction}	&\multicolumn{3}{c}{Knowledge Graph}\\
		\cline{2-7}
		&\#user	&\#item	&\#rating 	&\#entity	&\#relation	&\#triplet	\\
		\hline
		Last.FM		&1,872	&3,846	&42,346		&9,366	&60	&15,518	\\
		Movie-1M	&6,036	&2,347	&753,772	&6,729	&7	&20,195	\\
		%		outfit	&114,737	&30,040	&1,781,093	&0.0517$ \% $	&59,156	&51	&279,155	&0.00798$ \% $\\
		\hline
	\end{tabular}
    }
\end{table}

\begin{table*}[htp]
	\centering
	\caption{Overall comparison of performance}
	\label{table:overall}
    \resizebox{0.95\textwidth}{!}{
	\begin{tabular}{ c|c|ccc|ccc|ccc|ccc }
		\toprule[1.2pt]
		\multirow{2}{*}{Dateset} & \multirow{2}{*}{Model}	&\multicolumn{3}{c}{\itshape metric@1 $ (\%) $}	&\multicolumn{3}{c}{\itshape metric@5 $ (\%) $}	&\multicolumn{3}{c}{\itshape metric@10 $ (\%) $}	&\multicolumn{3}{c}{\itshape metric@20 $ (\%) $}\\
		\cline{3-14}
		& & recall & NDCG	& HR	& recall & NDCG	& HR	& recall & NDCG & HR	& recall & NDCG & HR \\
		\hline
		\multirow{6}{*}{Last}	
		&FM
		&1.93	&4.40	&4.40
		&5.33	&4.67	&12.80
		&8.83	&6.07	&19.40
		&14.02	&7.72	&28.01\\
		
		&NFM
		&1.50	&3.90	&3.90
		&5.95	&4.80	&13.20
		&9.52	&6.26	&21.10
		&14.97	&8.05	&29.90\\
		
		&CKE
		&4.43	&10.31	&10.31
		&13.06	&11.26	&26.58
		&18.85	&13.62	&35.02
		&26.95	&16.25	&46.11\\
		
		&KGAT
		&2.42	&5.67	&5.67
		&7.86	&9.49	&16.76
		&12.56	&12.58	&25.92
		&20.59	&16.71	&37.67\\
		
		&KGIN
		&\underline{6.06}	&\underline{13.98}	&\underline{13.98}
		&\underline{17.42}&\underline{15.24}
		&\underline{35.92}
		&\underline{24.96}	&\underline{18.32}&\underline{47.07}
		&\underline{35.49}	&\underline{21.69}&\underline{59.07}\\
		
		&proposed
		&\textbf{6.79} &\textbf{16.34}	&\textbf{16.34}
		&\textbf{20.15}	&\textbf{17.62}	&\textbf{39.84}
		&\textbf{28.05}	&\textbf{20.85}	&\textbf{50.69}
		&\textbf{38.78}	&\textbf{23.02}	&\textbf{61.84}\\
		\cline{2-14}
		&Improv.	
		&+12.05$ \% $	&+14.44$ \% $	&+14.44$ \% $	
		&+15.67$ \% $	&+15.62$ \% $	&+10.91$ \% $
		&+12.38$ \% $	&+13.81$ \% $	&+7.69$ \% $
		&+9.27$ \% $	&+6.13$ \% $	&+4.69$ \% $\\
		\hline

		\multirow{6}{*}{ML}	&FM
		&3.53	&32.70	&32.70
		&11.65	&27.57	&64.30
		&19.41	&26.88	&76.50
		&29.11	&27.98	&85.10\\
		
		&NFM
		&2.98	&27.70	&27.70
		&11.62	&24.88	&62.40
		&17.88	&23.80	&74.40
		&27.59	&24.84	&84.30\\
		
		&CKE
		&3.85	&\underline{33.54}	&\underline{33.54}
		&13.62	&\underline{28.78}	&\underline{66.65}
		&21.19	&\underline{27.89}	&\underline{78.29}
		&31.30	&\underline{29.18}	&\underline{86.51}\\
		%		Ripplenet	&0.0068	&0.0028	&0.0008	&0.0163	&0.1917	&0.1589	&0.1044	&0.7357\\
		&KGAT
		&2.63	&23.15	&23.15
		&10.03	&20.68	&57.01
		&16.98	&21.09	&71.59
		&26.37	&23.05	&82.15\\
		
		&KGIN
		&\textbf{4.69}	&11.99	&11.99
		&\textbf{15.14}	&12.92	&31.22
		&\underline{22.66}	&15.95	&43.22
		&\underline{31.50}	&19.35	&53.22\\
		
		&proposed
		&\underline{4.29}	&\textbf{36.74}	&\textbf{36.74}
		&\underline{15.01}	&\textbf{31.38}	&\textbf{70.13}
		&\textbf{23.29}	&\textbf{30.53}	&\textbf{80.55}
		&\textbf{34.12}	&\textbf{31.92}	&\textbf{88.34}\\
		\cline{2-14}
		&Improv.	
		&-8.69$ \% $	&+9.54$ \% $	&+9.54$ \% $	
		&-0.85$ \% $	&+9.03$ \% $	&+5.22$ \% $
		&+2.78$ \% $		&+9.47$ \% $	&+2.89$ \% $
		&+8.31$ \% $		&+9.39$ \% $	&+2.12$ \% $\\
		\bottomrule[1.2pt]
	\end{tabular}
    }
\end{table*}

In this section, we conduct empirical studies to demonstrate the effectiveness of proposed VRKG4Rec. The results of the experiments answer the following research questions:
\begin{itemize}
	\item \textbf{RQ1:} How does VRKG4Rec perform, comparing with the state-of-the-art KG-based approaches?
	\item \textbf{RQ2:} How does the key component, constructing VRKGs, contribute to model performance?
	\item \textbf{RQ3:} How does hyper-parameters (e.g.the number of iterations and the depth of GNN layer) affect model performance?
	\item \textbf{RQ4:} How does VRKG4Rec explore user preference and give an intuitive explainability?
\end{itemize}

\subsection{Experiment settings}
\subsubsection{Datasets}
We conduct our experiments on two public datasets: Last.FM\footnote{https://grouplens.org/datasets/hetrec-2011/} and MovieLens-1M\footnote{https://grouplens.org/datasets/movielens/1m/}, which is widely adopted in recommendation problems. Last.FM is built from Last.fm online music system. It contains music listening records of two thousand users. MovieLens-1M is a movie recommendation dataset, which consists of approximately 1 million rating data on the Movie-Lens website.
\par
Following the prior studies~\cite{wang:2019:KDD:KGNN-LS, wang:2019:WWW:MKR}, we convert the record data into user-item pair to serve as positive interaction data and we use Microsoft Satori to construct the knowledge graph for each dataset by matching the item IDs with the head of all triples. The basic statistics of the two datasets are presented in Table~\ref{table:dataset parameter}.
\par
For each dataset, we randomly select 80$ \% $ of positive interaction data for training, and treat the remaining as the test set.

\subsubsection{Competitors} We compare our proposed VRKG4Rec with the following algorithms that also exploit knowledge graph for recommendation.
\begin{itemize}
	\item \textbf{FM}~\cite{rendle:2011:SIGIR:FM} is a factorization model, which considers second-order feature interactions between items and a KG.
	\item \textbf{NFM}~\cite{he:2017:SIGIR:NFM} is a state-of-the-art factorization model, which modifying FM by using a neural network.
	\item \textbf{CKE}~\cite{zhang:2016:KDD:CKE} is an embedding-based method, which incorporates knowledge into the MF framework.
	\item \textbf{KGAT}~\cite{Wang:2019:KDD:KGAT} designs the knowledge graph attention network to model high-order conductivities in a KG.
	\item \textbf{KGIN}~\cite{wang:2021:WWW:KGIN} applies a GNN to explore user-item interactions by using auxiliary item knowledge.
\end{itemize}

\subsubsection{Evaluation metrics}
Following previous study KGIN~\cite{wang:2021:WWW:KGIN}, we conduct all-ranking strategy~\cite{walid:2020:KDD} in evaluation. Specifically, For each user in test set, We regard items that the user has interacted as positive items and treat the rest as testing items. We predict the ratings of all these testing items and select the items with highest ratings to form the recommendation list. To evaluate \textit{Top-n} recommendation performance. We adopt widely-used evaluation protocols: Recall, NDCG, Precision and HR. Without specification, we set the length of recommendation list as 20. The result is reported as the average metrics for all users in test set.

\subsubsection{Parameter setting}
We implement our proposed model in PyTorch. We fix the size of ID embeddings {\itshape d} as 64, the optimizer as Adam~\cite{diederik:2015:ICLR:Adam}, and the batch size as 1024. Without specification, we set the number of VRKGs $ K $ as 3. Meanwhile, we fix the number of iteration scales $ Q $ as 3 and GNN layers $ L $ as 2. Besides, the learning rate is set as $ l_r = 10^{-4} $ and the $ L_2 $ penalty is set as $ 10^{-5} $. We make a test per 10 epochs and the total epoch number is set as 1000.

%
%\begin{figure*}[t]
%	\centering
%	\subfigure[Performance of VRKG4Rec with different iteration $ Q $ and layer $ L $ for Last-FM]{
%		\begin{minipage}[t]{0.96\linewidth}
%			\centering
%			\includegraphics[width=0.32\linewidth]{figure/last1.png}
%			\includegraphics[width=0.32\linewidth]{figure/last2.png}
%			\includegraphics[width=0.32\linewidth]{figure/last3.png}
%			%\caption{fig1}
%		\end{minipage}%
%	}%
%	\hfill
%	\subfigure[Performance of VRKG4Rec with different iteration Q and Layer L for MovieLens-1M]{
%		\begin{minipage}[t]{0.96\linewidth}
%			\centering
%			\includegraphics[width=0.32\linewidth]{figure/movie1.png}
%			\includegraphics[width=0.32\linewidth]{figure/movie2.png}
%			\includegraphics[width=0.32\linewidth]{figure/movie3.png}
%			%\caption{fig1}
%		\end{minipage}%
%	}%	
%	\centering
%	\caption{Impact of iteration Q and Layer L}
%	\vspace{-0.2cm}
%	\label{fig:hyper}
%\end{figure*}

\subsection{Performance Comparison(RQ1)}
We vary the length of the recommendation list in $ \{1, 5 , 10, 20\} $ to conduct experiments. Table~\ref{table:overall} presents the overall performance comparison between our VRKG4Rec and the competitors, where the best results in every column are boldfaced and the second are underlined. It can be observed that our VRKG4Rec outperforms the competitors on both two datasets in most of cases. We conduct some analysis and discussions as follows:
\par
We attribute the better performance of VRKG4Rec to our conversion of many relations of a knowledge graph into a few virtual relations. Such virtual relations can not only find out the relevance relations in KG, but also help encoding those relational knowledge more beneficial to the downstream recommendation task. Besides, our LWS for encoding an item merely from its neighbor embeddings. Such an encoding mechanism focuses on converting local relational knowledge into neighboring item encodings, as it tries to ensure relational knowledge-connected entities with closer distances in an embedding space.
\par
The FM and NFM perform poorly on both datasets, because factorization model is unable to fully exploit the relations of KG to facilitate item representation. The embedding-based method CKE utilizes TransR to encode the first-order entity and relation information of a KG for entity representation learning; While the KGAT and KGIN design graph neural models so as to capture high-order neighbors' information and relational knowledge. This operation, however, can be either constructive or destructive to recommendation.
\par
It can be observed that the KGAT performs rather worse on both two dataset. We note that the KG for Last-FM dataset contains 60 relation types and about 155K knowledge triplets; While the Movielens-1M contains only 15 relation types and about 200K knowledge triplets. Furthermore, both items and entities are fewer in the Movielens-1M dataset. The worse performance of KGAT on the Last-FM dataset suggest that in a sparse KG but with many relation type, it is not wise to consider all relations in item encoding. And on relatively dense dataset, it may lead to overfit to discriminate the attention of all relations. We note that our VRKG4Rec is robust to different KG scales for its conversion of virtual relations.
\par
KGIN and CKE have opposite performance on the two dataset, where KGIN achieves better performance than CKE on Last.FM and worse than CKE in terms of most metrics on MovieLens-1M. It is because that the KG for MovieLens-1M mainly consists of one-hop knowledge triples and TransR in CKE can better learn this connectivity while KGIN may introduce noise information during high-order propagation.

\begin{figure}[t]
	\centering
	\subfigure[Last.FM]{
		\begin{minipage}[t]{0.5\linewidth}
			\centering
			\includegraphics[width=1\textwidth]{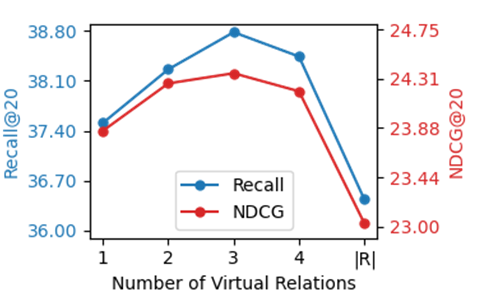}
			%\caption{fig1}
		\end{minipage}%
	}%
	\subfigure[Moive-1M]{
		\begin{minipage}[t]{0.5\linewidth}
			\centering
			\includegraphics[width=0.99\textwidth]{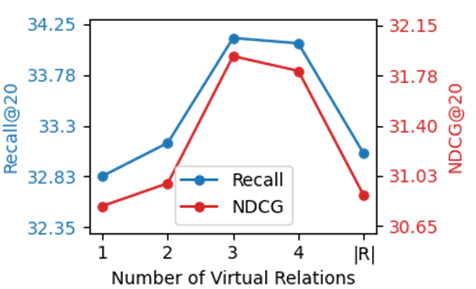}
			%\caption{fig2}
		\end{minipage}%
	}%
	\centering
	\caption{Impact of VRKGs}
	\label{Fig:zhexian}
\end{figure}

\subsection{Impact of VRKG(RQ2)}\label{sec:ablation}
In order to evaluate the effectiveness of VRKG construction and virtual relation clustering for item encoding, we consider two KG variants: (1)Variant-1 doesn't construct VRKG but directly apply LWS on the KG without discriminating relation types; (2)Variant-2 discards virtual relation clustering but construct VRKGs based on origin KG. For each relation type, we construct one relational knowledge graph, on which the LWS is applied. To further analyze the impact of virtual relation number, we set the value of $ K $ in range of $ \{2, 3, 4\} $. We report the performance of \textit{Recall@20} and \textit{NDCG@20} in Fig. ~\ref{Fig:zhexian}, where variant-1 is corresponding to $ K=1 $ while variant-2 is to $ K=|\mathcal{R}| $. Note that $ |\mathcal{R}| $ is the number of relations in origin KG. From Fig. ~\ref{Fig:zhexian}, we can observe that:

On the one hand, the worse performance of variant-1 indicates the effectiveness of construing VRKGs. We attribute its superiority to two reasons. (1) Variant-1 treats knowledge graph as a homogeneous graph, ignoring the rich relational information. (2) Furthermore, Variant-1 aggregates the information from all neighbors into one embedding, which distorts the semantic independence and integrity of item attributes. On the other hand, the performance of variant-2 goes down dramatically, indicating that not every relation type for item encoding is constructive to recommendation.

We notice that discarding virtual relation clustering has greater impact on Last-FM than Movie-1M dataset. The possible reason is that the KG in Movie-1M doesn't have that many relation types like that in Last-FM and each type of relations is relatively independent for item representation. This again validates the necessity of constructing virtual relations especially on the KG with a large number of edges.

On both two datasets, the curves of Recall and NDCG first increase and then decline after reaching peak performance when $ K=3 $. It indicates that clustering the type of relations will help to capture the relevance of relations. However, with the number of virtual relations increasing, some highly relevant entities will be separated into different VRKGs in item encoding, which may prevent useful information passing.

\begin{figure}
	\centering
	\subfigure[Last-FM]{
		\begin{minipage}[t]{0.47\linewidth}
			\centering
			\includegraphics[width=1.5in]{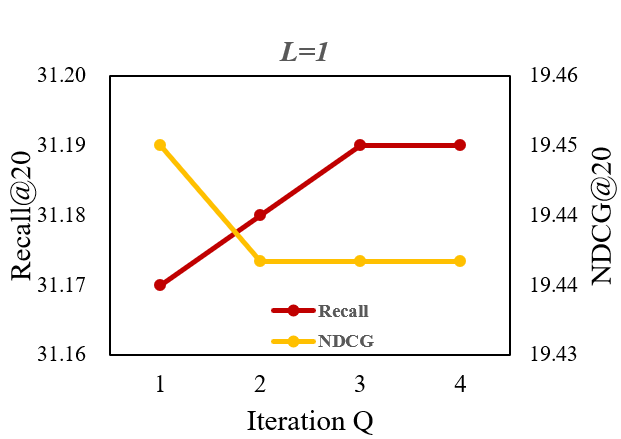}\\
			\vspace{0.02cm}
			\includegraphics[width=1.5in]{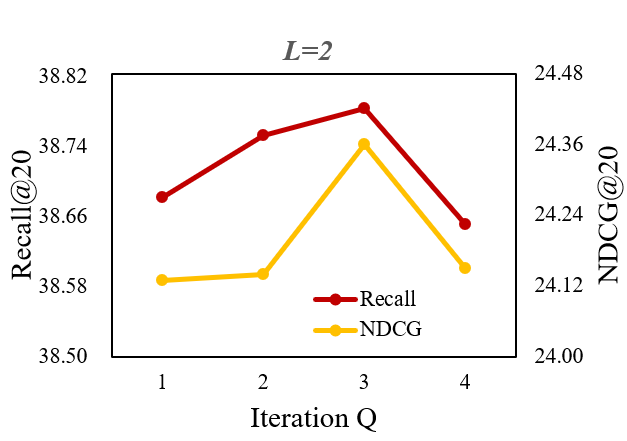}\\
			\vspace{0.02cm}
			\includegraphics[width=1.5in]{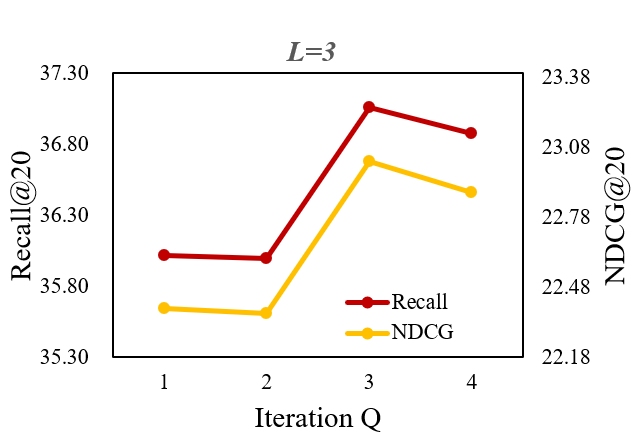}\\
			\vspace{0.02cm}
		\end{minipage}%
	}%
	\subfigure[Movie-1M]{
		\begin{minipage}[t]{0.47\linewidth}
			\centering
			\includegraphics[width=1.5in]{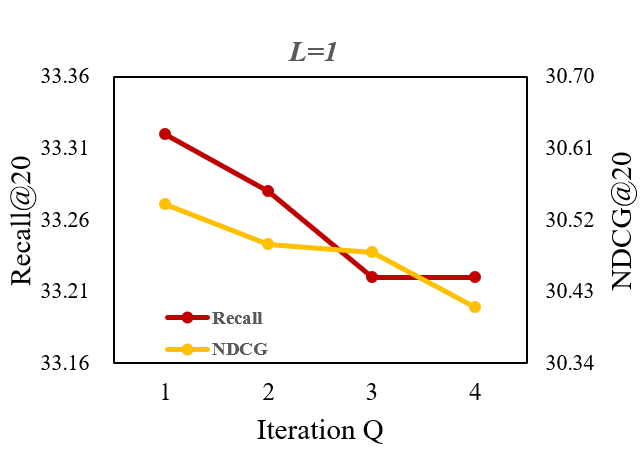}\\
			\vspace{0.02cm}
			\includegraphics[width=1.5in]{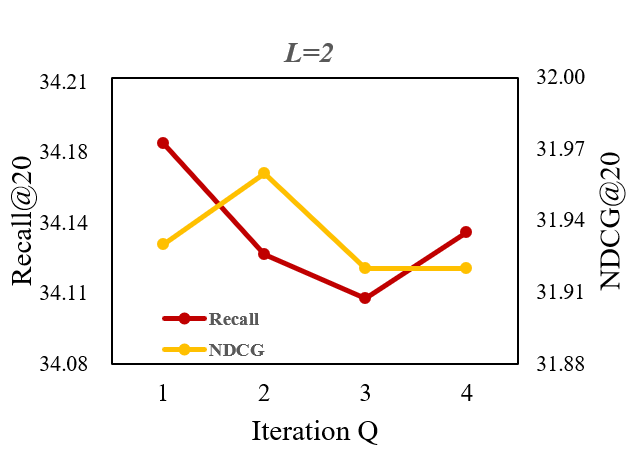}\\
			\vspace{0.02cm}
			\includegraphics[width=1.5in]{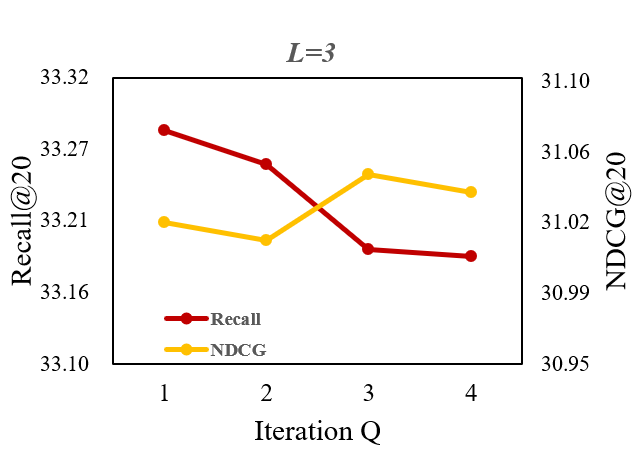}\\
			\vspace{0.02cm}
		\end{minipage}%
	}%
	\centering
	\caption{Impact of iteration Q and Layer L}
	\vspace{-0.2cm}
	\label{fig:hyper}
\end{figure}

\subsection{Parameter Sensitivity(RQ3)}
In this section, we analyze the impact of the number of LWS iterations $ Q $ and the number of the GNN layers $ L $ on the performance of VRKG4Rec model. We perform the experiment on Last-FM and MovieLens-1M dataset while setting the number of virtual relations $ K $ as 3. Fig.~\ref{fig:hyper} illustrates the performance of \textit{Recall@20} and \textit{NDCG@20} on both two datasets.

In Fig.~\ref{fig:hyper}(a) and Fig.~\ref{fig:hyper}(b), we fix the number of GNN layer $ L $ and change the iteration number $ Q $ in range of $ \{1, 2, 3, 4\} $. We notice that the performance curve will also first increase and then decline on Last-FM dataset. The result indicates that with $ Q $ increase, the item will closer to the similar neighbor in embedding space, which will benefit item representation in recommendation task. Nevertheless, with $ Q $ getting bigger, the embeddings of nodes will be too close to discriminate, thus doing harm to model performance. Furthermore, the performance shows a downward trend when iteration $ Q $ increase on MovieLens-1M, for the reason that the KG in MovieLens-1M has richer triples and less entities than that in Last-FM and the denser connection make it easier for embedding to be over-smoothing.
\par
If we fix the number of iteration $ Q $ and increase $ L $ in range of $ \{1, 2, 3\} $, the performance will first goes up and then falls down. Specifically, model with $ L=2 $ outperforms that with $ L=1 $, as the latter only explores the first-order connectivity in KG, leaving the high-order semantic information untouched. However, the performance get worse when $ L=3 $, this could be that irrelevant or noise information is generated during long-range propagation.
\par
Summarily, with the layer $ L $ increasing, the model performance will first improve significantly, and then degrade across-the-board regardless of the iteration number $ Q $. We can draw the conclusion that the GNN layser $ L $ plays a decisive part in model performance. Because the number of layer $ L $ determines the depth of receptive field. Stacking $ L $ layers will propagate information from \textit{L-hop} neighbors into item embedding. While iteration $ Q $ fine-tunes the information from  \textit{1-hop} neighbors to control the smoothness of the item embedding and its neighbor embeddings.

\begin{figure}[t]
	\centering
	\includegraphics[scale=0.6]{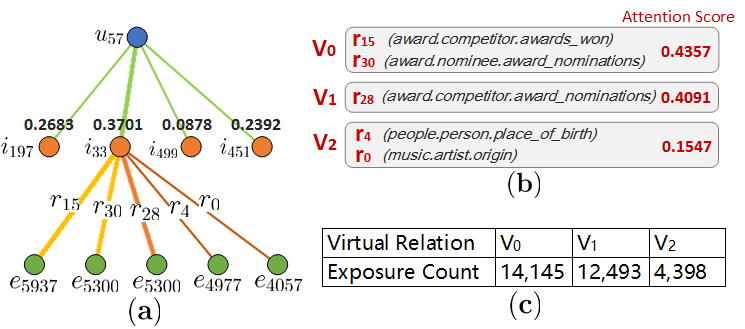}
	\caption{Case study for from Last-FM dataset. (a) shows the preference of user $ u_{57} $ on the interacted items and the clustered relations. Best viewed in color. (b) shows semantics of the relations in case study and the corresponding virtual relations. (c) shows exposure count of virtual relations.}
	\label{Fig:casestudy}
\end{figure}

\subsection{Case study(RQ4)}
In case study, We take user $ u_{57} $ and item $ i_{33} $ from Last-FM dataset as an example to give an intuitive explainability. We further exhibit the exposure count of virtual relations to validate effectiveness of alleviating the long-tail issue. Fig.~\ref{Fig:casestudy} presents the visualization result. Specifically, Fig.~\ref{Fig:casestudy}(a) shows the preference of user $ u_{57} $ on the interacted items and clustered relations, where the blue, orange, green node denotes user, item and entity respectively. Fig.~\ref{Fig:casestudy}(b) shows the semantics of the relations in case study and the corresponding virtual relations. Fig.~\ref{Fig:casestudy}(c) show the exposure count of the virtual relations. 

Fig.~\ref{Fig:casestudy}(a) exhibits that user's preference for the interacted items is different and $ i_{33} $ is more important to $ u_{57} $'s preference modeling for it enjoys a higher weight. From Fig.~\ref{Fig:casestudy}(a) and Fig.~\ref{Fig:casestudy}(b), We observe that $ r_{15} $ and $ r_{30} $ are replaced by virtual relation $ v_0 $, for they are highly relevant and both related to \textit{award}. And clearly, relations aligned with virtual relation $ v_2 $ is more related to \textit{place}.   The attention score in Fig.~\ref{Fig:casestudy}(b) is the weight in item embedding fusion step (equation (18)). It denotes user's attention on the virtual relations. From the attention scores, we can infer that user $ u_{57} $ cares more about the \textit{award} than \textit{place} information when choosing an item. Thus, it can be explained that \textit{User $ u_{57} $ selects item $ i_{33} $ because it is involved with the award $ u_{57} $ cares for. }  Fig.~\ref{Fig:casestudy}(c) shows that the exposure count of the virtual relations is comparable and we can conclude the relation long-tail issue is alleviated by introducing virtual relations.

\section{Related Work}\label{sec:related work}
As noted in the introduction, the mostly related work on RS with knowledge graph can be divided into three categories:
\subsection{Embedding-based Methods}
Embedding-based methods~\cite{wang:2018:WWW:DKN, zhang:2016:KDD:CKE, Palumbo:2017:recsys:entity2rec, cao:2019:WWW:KTUP, ai:2018:algorithms:ECFKG, zhang:2018:arxiv:CFKG, wang:2019:WWW:MKR} mainly take advantage of the techniques in knowledge graph embedding (eg. transE~\cite{bordes:2013:transE}) to get item embedding which is used as pre-trained knowledge in subsequent recommendation task. Wang et al.~\cite{wang:2018:WWW:DKN} design a deep knowledge-aware network to mine latent knowledge-level relations in between news for generating news' representations. Zhang et al.~\cite{zhang:2016:KDD:CKE} hire TransE on knowledge graph and feed the learned embedding into matrix factorization(MF). Wang et al.~\cite{wang:2019:WWW:MKR} propose a multi-task feature learning model based on a deep semantic matching framework to learn item and entity embedding by a cross\&compress unit. Although these methods absorb the facts in KG as side information, they mainly focus on the first-order connectivity and ignore the global relationships of entities.

\subsection{Path-based Methods}
Path-based methods ~\cite{catherine:2016:recsys, ma:2019:WWW, fan:2019:KDD:metapath, zhao:2017:KDD:meta, hu:2018:KDD:leveraging} sample the KG to explore a path between a target user and a candidate item for discovering some latent relations in between them. For example, Ma et al.~\cite{ma:2019:WWW} propose a joint learning framework to leverage associations between items in KG for delivering explainable recommendation. Fan et al.~\cite{fan:2019:KDD:metapath} present a metapath-guided embedding method to take full advantage of side information for well representing user intents. Zhao et al.~\cite{zhao:2017:KDD:meta} characterize the rich semantics of the heterogeneous types of entities by combining meta-graph with matrix factorization and factorization machine. Hu et al.~\cite{hu:2018:KDD:leveraging} design a co-attention mechanism for mining the mutual effect between different metapaths. However, this kind of methods suffers from two main problems: (1)It needs professional knowledge and is weak in transfer ability. (2)It is labor-intensive and time-consuming when applying on KG of large scale.

\subsection{GNN-based Methods}
GNN-based methods ~\cite{wang:2018:CIKM:ripplenet, wang:2019:WWW:KGCN, wang:2019:KDD:KGNN-LS, Wang:2019:KDD:KGAT, tang:2019:KDD:AKUPM, wang:2020:SIGIR:CKAN, cao:2021:SIGIR:dekr, wang:2021:WWW:KGIN} propagate information through the topology of knowledge graph and aggregate the information from the neighbor node to update the representation of the center nodes. Wang et al.~\cite{wang:2018:CIKM:ripplenet} propose an end-to-end framework Ripplenet to iteratively extend user's potential interests along connections in a KG in order to stimulating the propagation of user's preferences. KGAT~\cite{Wang:2019:KDD:KGAT} models the high-order interactions in KG explicitly based on attention mechanism and applies the same attention network to learn user preference. Cao et al.~\cite{cao:2021:SIGIR:dekr} incorporate text-based method and knowledge graph-based method to make full use of description information. Tang et al.~\cite{tang:2019:KDD:AKUPM} devise AKUPM which leverages a self-attention network to weight the importance of intra-entity-interaction and inter-entity-interaction for learning user representation. KGIN~\cite{wang:2021:WWW:KGIN} as a state-of-the-art GNN-based recommendation model, encodes relation embedding into representation to preserve the semantic of long-range connectivity in KG.

GNN-based methods is really popular with researchers nowadays due to the excellent and promising performance. However, to the best of our knowledge, current GNN-based methods directly apply GNN scheme on the origin KG which contains relations with similar function and may not suitable for recommendation tasks. Additionally, existing GNN-based methods aggregate all relational information into one embedding, which, we argue, will damage semantic information. In this paper, we propose to construct virtual relational knowledge graphs to preserve the semantic integrity and devise a local weighted smoothing (LWS) aiming at collecting side information from VRKGs to facilitate item representation for better recommendation performance. 

\section{Conclusion}\label{sec:conlusion}
Although knowledge graph is helpful for enriching item and user representation, we argue that it is not wise to directly exploit all relations of an origin KG without considering a particular downstream task. In this work, we have proposed a VRKG4Rec model to first construct VRKGs to learn a kind of virtual relational knowledge for item encoding. We have designed the LWS, a new graph neural model, for node encoding in a graph, which has been applied in each VRKG for item encoding. A fusion mechanism is used to learn final item representation. Experiments on two datasets have shown that the proposed VRKG4Rec outperforms the state-of-the-art methods. It is worth mentioning that we set the number of VRKGs manually in our model and it is labor-intensive and time-consuming to find the optimum number. Our future work shall investigate how to learn the optimal VRKG number automatically.

\section*{Acknowledgements}
This work is supported in part by National Natural Science Foundation of China (Grant No: 62172167). We also want to use our VRKG4Rec model on MindSpore2 \footnote{http://www.mindspore.cn/}, which is a new deep learning computing framework. These problems are left for future work.

\bibliographystyle{ACM-Reference-Format}
\bibliography{reference}
\end{document}